%% file: main.tex
\documentclass[10pt,conference]{IEEEtran}

\input{preamble}   
    
\begin{document}

\title{Does the duration of rapid release cycles affect the bug handling activity?}

\author{\IEEEauthorblockN{Thorn Jansen\IEEEauthorrefmark{1}, Zeinab Abou Khalil\IEEEauthorrefmark{2}, Eleni Constantinou\IEEEauthorrefmark{1}, and Tom Mens\IEEEauthorrefmark{2}}

\IEEEauthorblockA{\IEEEauthorrefmark{1} Eindhoven University of Technology,
Netherlands, 
Email: t.k.h.g.jansen@student.tue.nl, e.constantinou@tue.nl}
\IEEEauthorblockA{\IEEEauthorrefmark{2} University of Mons, Belgium,
Email: \{zeinab.aboukhalil,tom.mens\}@umons.ac.be}
}

%
%
%
%
%



\maketitle

\begin{abstract}
\input{sections/abstract}
\end{abstract}


%
%
%
%
%





\section{Introduction}\label{sec:intro}
\input{sections/introduction}

\section{Methodology}\label{sec:methodology}
\input{sections/methodology}

\section{Results}\label{sec:results}
\input{sections/results}

\section{Discussion}\label{sec:discussion}
\input{sections/discussion}

\section{Threats to Validity}\label{sec:threats-validity}
\input{sections/threaths-validity}

\section{Related Work}\label{sec:rel-work}

\input{sections/related-work}

\section{Conclusions}\label{sec:conclusion}
\input{sections/conclusion}
\bibliographystyle{IEEEtran}

\bibliography{references}

\end{document}

%% file: preamble.tex
\usepackage{amsmath,amssymb,amsfonts}
\usepackage{graphicx}
\usepackage{subfig}
\usepackage[dvipsnames]{xcolor}
\usepackage{comment}
\usepackage{hyperref}
\usepackage{url}

\usepackage{cite}
\usepackage{algorithmic}
\usepackage{textcomp}
\usepackage{xcolor}

\usepackage{verbatim}
\usepackage{mdframed}

\usepackage{xspace}
\usepackage[utf8]{inputenc}

\usepackage{tcolorbox}

\usepackage{soul}

\newcommand{\etal}{et al.\xspace}
\newcommand{\ie}{i.e.,\xspace}

\newcommand{\fig}[1]{Fig.~\ref{#1}}

\newcommand{\sect}[1]{Section~\ref{#1}}

\usepackage{url}

\usepackage{subfig}

\usepackage[]{todonotes}

\newtcolorbox{mybox}{colback=gray!30,
boxrule=0pt,arc=0pt,boxsep=2pt,left=2pt,right=2pt,leftrule=1pt}

\newmdtheoremenv[]{summary}{Summary\ignorespaces}

\usepackage{balance}
\usepackage{changepage}
\usepackage{framed}
\makeatletter
 {\par\unskip\endMakeFramed}
\makeatother

\definecolor{formalshade}{rgb}{0.93,0.93,0.93}

\definecolor{darkblue}{rgb}{0.2, 0.2, 0.2}

\newenvironment{formal}{%
  \def\FrameCommand{%
    \hspace{1pt}%
    {\color{darkblue}\vrule width 2pt}%
    {\color{formalshade}\vrule width 4pt}%
    \colorbox{formalshade}%
  }%
  \MakeFramed{\advance\hsize-\width\FrameRestore}%
  \noindent\hspace{-1pt}
  \begin{adjustwidth}{}{7pt}%
  \vspace{2pt}\vspace{2pt}%
}
{%
  \vspace{3pt}\end{adjustwidth}\endMakeFramed%
}

\newcounter{resultcounter}

\newcounter{patterncounter}

\usepackage{longtable,booktabs}
\usepackage{pdfpages,amsmath,amssymb}
\usepackage{amssymb}
\usepackage{mathtools, nccmath}

\usepackage{cleveref}

\usepackage{float}
\DeclarePairedDelimiterX{\set}[1]{\{}{\}}{\setargs{#1}}
\NewDocumentCommand{\setargs}{>{\SplitArgument{1}{;}}m}
{\setargsaux#1}
\NewDocumentCommand{\setargsaux}{mm}
{\IfNoValueTF{#2}{#1} {#1\,\delimsize|\,\mathopen{}#2}}

\DeclarePairedDelimiter\abs{\lvert}{\rvert} 
\DeclarePairedDelimiter\norm{\lVert}{\rVert}
\makeatletter
\let\oldabs\abs
\def\abs{\@ifstar{\oldabs}{\oldabs*}}
\let\oldnorm\norm
\def\norm{\@ifstar{\oldnorm}{\oldnorm*}}
\makeatother
\usepackage{tabularx}
\usepackage{tcolorbox}

\newcommand{\linebreakand}{%
  \end{@IEEEauthorhalign}
  \hfill\mbox{}\par
  \mbox{}\hfill\begin{@IEEEauthorhalign}
}

%% file: sections/abstract.tex
Software projects are regularly updated with new functionality and bug fixes through so-called releases. In recent years, many software projects have been shifting to shorter release cycles and this can affect the bug handling activity. 
Past research has focused on the impact of switching from traditional to rapid release cycles with respect to bug handling activity, but the effect of the rapid release cycle duration has not yet been studied. 
We empirically investigate releases of 420 open source projects with rapid release cycles to understand the effect of variable and rapid release cycle durations on bug handling activity.
We group the releases of these projects into five categories of release cycle durations.
For each project, we investigate how the sequence of releases is related to bug handling activity metrics
and we study the effect of the variability of cycle durations on bug fixing. 
Our results did not reveal any statistically significant difference for the studied bug handling activity metrics in the presence of variable rapid release cycle durations. This suggests that the duration of fast release cycles does not seem to impact bug handling activity.

%% file: sections/introduction.tex
Many contributors to open source software (OSS) projects work on a volunteer basis~\cite{crowston2007self}, making it challenging to attract and retain developers and users~\cite{subramaniam2009determinants}. To keep users interested, it is important to release new functionalities and avoid having bugs that detriment users’ overall experience with the software. New versions of the software are delivered periodically in so-called \textit{releases} that provide new user functionality and aim to resolve as many bugs as possible~\cite{adams2016modern}.

Large organizations like Google and Facebook, and foundations like Mozilla and Eclipse have migrated from traditional long release cycles to more rapid release cycles \cite{mantyla2015rapid} to provide users with bug fixes and new functionality more frequently. 
Regardless of the release cycle duration, developers seek to deliver software with as few bugs as possible.  However, the adoption of rapid releases may result in less time for the community to address bugs~\cite{aboukhaliljss2020}.

While switching to rapid release cycles has shown to be beneficial for mature projects with large developer communities such as Eclipse~\cite{abou2019longitudinal,aboukhaliljss2020} and Firefox \cite{mantyla2015rapid,khomh2015understanding}, it remains unknown whether and how such a switch impacts projects with different characteristics. The optimal release cycle duration for different projects may differ depending on the size of the developer community, project size, development processes used and level of automation. 
Moreover, the rapid release cycle duration differs across projects (e.g., Eclipse follows a 13-week cycle while Firefox follows a 4-week cycle), and even within the same project the cycle duration may fluctuate. It therefore becomes important to investigate how varying release cycles within a project can affect its software development processes. 
Nonetheless, projects that switch to rapid cycles of fixed or variable duration aim to sustain or improve the bug handling performance of their developer community. 

To understand the effect of rapid release engineering on bug handling, we carry out an empirical investigation. More specifically, we focus on the effect of the duration of rapid release cycles on the fixing, triaging, and survival of bugs. 
To do so, we rely on the RapidRelease dataset~\cite{joshi2019rapidrelease} that contains 994 GitHub projects with over 3 million issue reports and with an average release cycle duration between 5 and 35 days. We group releases in five categories of release duration, and treat the releases of each project as a sequence of release categories. This allows us to use the Gandhi-Washington Method (GWM) implemented by \textit{Nayebi et al.}~\cite{nayebi2019mining}. The GWM tool allows us to test if there is a statistically significant difference in a given metric w.r.t different release cycle durations.
Our investigation is guided by four research questions:

$RQ_1:$ \textit{How does release cycle duration affect the bug fixing process?}
This question focuses on bug fixing, which is the main concern of the bug handling process. To investigate how bug fixing is affected by the release cycle duration, we operationalize bug fixing in terms of fixing duration and fixing ratio and apply GWM on these metrics.

$RQ_2:$ \textit{How does release cycle length affect bug survival over multiple releases?}
The longer a bug is present, the more costly it becomes to fix it~\cite{boehm1988understanding}.  
Therefore, in presence of rapid release cycles it is relevant to examine the likelihood for a bug to survive over releases. 
We hypothesize that with shorter release cycles  developers have less time for new feature implementation and bug fixing, making it more likely for bugs to ``survive" over releases until they are handled.


$RQ_3:$ \textit{How does release cycle duration affect bug triaging?}

The step of assigning bug reports to developers is called bug triaging~\cite{hu2014effective}. 
Bug triaging can affect the bug handling activity, e.g., multiple re-assignments impact how long a bug stays unfixed~\cite{saha2015understanding,saha2014empirical}. 
Therefore, we investigate the effect of the release cycle duration on bug triaging, operationalised in terms of triaging duration and ratio.
A short triaging duration and a high triaging ratio would benefit the overall bug handling performance.


$RQ_4:$ \textit{How do variations in the release cycle duration of a project affect the bug handling process?}
Although the average or median release cycle duration describes the general behavior of the release engineering within a software project, it does not explain any variance in the release cycle duration of a project. Two projects can have a similar average release cycle duration but a different sequence of release cycle durations. 
Alternating between different periods of time to publish a new release, can require additional time to carry out other development activities within a given release, thus leaving less time for bug handling compared to a sequence of releases of the same duration. 
Therefore, an interesting factor to evaluate is if and what effect the variability of release cycle durations within a project has on the bug handling activity.   


The remainder of the paper is organized as follows.  \sect{sec:methodology} describes the research methodology and data extraction process, while  \sect{sec:results} presents the quantitative results. \sect{sec:discussion} discusses our findings and \sect{sec:threats-validity} reports on the threats to validity.  \sect{sec:rel-work} discusses the related work. 
Finally, \sect{sec:conclusion} concludes and presents future research.

%% file: sections/methodology.tex
This section presents the selected dataset and data extraction process, the definitions of the metrics used, and the methodology used to address the research questions. 
The datasets and scripts used for the analysis are available in a replication package on Zenodo \cite{MSR2020}. 


\subsection{Datasets}





To perform our empirical analysis, a set of projects has to be chosen to analyze the relation between bug handling and release cycle duration.
We selected the 994 software repositories in the RapidRelease dataset~\cite{joshi2019rapidrelease,RapidReleaseGIT} 
  because this dataset explicitly contains repositories with a short duration between releases (5 to 35 days on average).
Secondly, the repositories combined have over three million issue reports. Although GitHub issues do not always represent actual bug reports, a very large sample of issue reports increases the probability of a large dataset of bug reports. 
The RapidRelease dataset contains information about the project releases, but this does not include information on the issues present, and event information like when developers are assigned to an issue, or when a commit is attached to an issue.
Therefore, based on the repository names provided in the dataset, we used the GitHub API to enhance the metadata of each project in three ways:

\begin{itemize}
\item Releases: we collect all project releases and their attributes, such as the date the release was published.
\item Issues: we fetch information on all issues (both open and closed), including information about the opening and closing date, the labels used by the reporter to categorize the issue, and whether a developer was assigned to the issue.
\item Issue events: for each issue, we collect the date when developers are assigned (if assignment took place). 
\end{itemize}





\begin{figure}[!h]
    \centering
    \includegraphics[width=0.6\columnwidth]{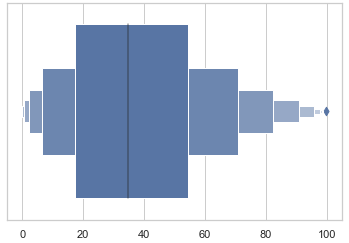}
    \caption{Boxen plot showing the distribution of the ratio of labeled issues per project.}
 
    \label{fig:issues-covered}
\end{figure}

As mentioned earlier, GitHub issues do not necessarily correspond to bug reports.
For example, they may be used to request new functionalities or to ask questions to developers~\cite{bissyande2013got}. 
To differentiate between bugs and other types of issues, we collected the labels of all issues of the selected repositories.
We found a total of 3,520,016 issues where 1,784,735 issues had at least one label out of a total of 5,532 unique labels.
However, labelled issues are not evenly distributed between the projects. The boxenplot \cite{boxenplot} in Figure~\ref{fig:issues-covered} provides an overview of the ratio of issues labeled per project. 

\begin{figure}[!h]
    \centering
    \includegraphics[width=0.6\columnwidth]{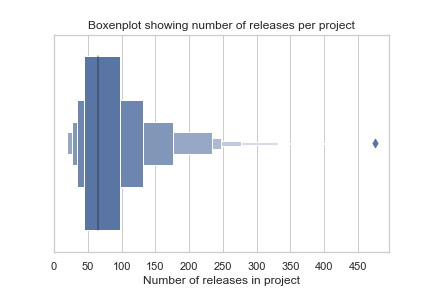}
    \caption{Boxen plot showing the distribution of the number of releases per project.} 
    
    \label{fig:num-release-project}
\end{figure}

Because projects with a low rate of labeled issues could bias the results, we decided to filter out projects that have labels in less than 40\% of their issues. 
This leaves 420 projects to be analyzed. Figure~\ref{fig:num-release-project} shows the distribution of the number of releases in each project of this filtered dataset and Table~\ref{tab:stats-num-releases} shows the general statistics about the number of releases of the 420 projects. As all projects have at least 15 releases with a median of 65 releases, we ensure a sufficient number of releases per project.

\begin{table}[h]
\begin{tabular}{l||l}
\textbf{Average number of releases} & 79.767 \\\hline
\textbf{Median number of releases} & 65 \\\hline
\textbf{Minimum number of releases} & 15 \\\hline
\textbf{Maximum number of releases} & 475 \\\hline
\textbf{Standard deviation number of releases} & 57.073 \\
\end{tabular}
\caption{Descriptive statistics of the number of releases in the 420 repositories analyzed.
\label{tab:stats-num-releases}}
\end{table}

Furthermore, Table~\ref{tab:stats-num-issues} contains general statistics of the number of issues present in the 420 projects. As all projects have at least 57 issues with a median of 3099.5 issues, we ensure a sufficient number of issues per project to investigate bugs handling activity. 

\begin{table}[h]
\begin{tabular}{l||l}
\textbf{Total number of issues} & 1,493,393 \\\hline
\textbf{Average number of issues} & 3555.698 \\\hline
\textbf{Median number of issues} & 3099.5 \\\hline
\textbf{Minimum number of issues} & 57 \\\hline
\textbf{Maximum number of issues} & 41,904 \\\hline
\textbf{Standard deviation number of issues} & 6289.336 \\
\end{tabular}
\caption{Descriptive statistics of the number of issues in the 420 repositories analyzed.
\label{tab:stats-num-issues}}
\end{table}


To determine which labels represent bugs, the four authors of this paper independently and manually classified which labels concern bugs; the inter-rater agreement (IRR) was measured using Cohen’s $\kappa$~\cite{mchugh2012interrater}. At least 2 authors categorized each label, and in case of disagreement, all authors discussed to reach a consensus.
On the first round of manual labeling, an agreement of $\kappa=0.52$ was obtained. Most of the disagreements were caused by a different perception between authors of labels containing the word ``\textit{triage}''. After closer inspection and discussion, the raters decided that this label did not concern bug triaging exclusively, and thus it was excluded from the set of bug-specific labels. After the second round of manual labeling, a perfect agreement of $\kappa=1$ was obtained.

The manual labeling resulted in a total of 183 labels representing bugs found in 327,579 labeled issues. 
Although the bug labels correspond to only 3\% of all unique labels, they correspond to 22\% of all labeled issues (327,579 out of 1,493,393).
This is caused by bug relating labels occurring often, e.g. ``\textit{bug}" is one of the commonly used bug-related labels in different repositories.
On the contrary, a large number of labels are only used a few times, but not representing a large number of issues, e.g., ``\textit{version/1.1}".

\subsection{Definition of metrics}


To evaluate the effect of the release cycle duration on the bug handling process, we need to compare the release information to bug information. To do so, for a given project $p$ we use the notation $B^p$ to refer to the set of all bugs reported for $p$. Similarly, $B_{open}^p$ refers to the subset of all open bugs, $B_{triage}^p$ to the subset of all triaged bugs, $B_{fix}^p$ to all fixed bugs and $B_{close}^p$ to all closed bugs. Let $R_p$  be the set of all project releases of $p$. Given a release $r \in R_p$, $B^{p:r}$ constrains the set $B^p$ to only those bugs targeting release $r$. For example, $B^{p:4.0}_{fix}$ contains all bugs fixed in release 4.0 of project $p$.

According to \textit{Liu et al.}~\cite{liu2016comparative}, a bug can be considered fixed if there is at least one commit attached to the issue and the issue has been closed on GitHub. Additionally, referencing an issue in a commit if the commits addresses the issue is the recommended GitHub behavior~\cite{kalliamvakou2014promises}. Additionally, \textit{Liu \etal}~\cite{liu2016comparative} define the bug fixing moment to be the point in time when the issue was closed, not when the commit was attached. This is important for defining the fixing duration. 
Moreover, a bug is triaged when a developer gets assigned to the issue representing the bug~\cite{abou2019longitudinal}. 
Therefore, we define the following auxiliary functions:
\begin{itemize}

    \item date: $R_p \rightarrow Date$  returns the publication timestamp of a given release of project $p$.
    \item earlier: $R_p \rightarrow \mathcal{P}(R_p)$ returns the set of earlier releases of a given release of project $p$.
    \item prev: $R_p \rightarrow R_p$ returns the release immediately preceding a given release of project $p$.
    \item $D_{\text{open}}: B^p \rightarrow Date$ returns the timestamp when the bug report was opened. 
    \item $D_{\text{triage}}: B^p \rightarrow Date$ returns the timestamp when the bug report was triaged. 
    \item $D_{\text{close}}: B^p \rightarrow Date$ returns the timestamp when the bug report was closed.
    \item $d(r) = [\text{date}(\text{prev}(r)), \text{date}(r)]$ is the time interval between a release $r\in R_p$ and the previous release.
    \item $B_\text{newlyClosed}^p(r) = \bigcup_{e \in earlier(r)} \{ b\in B^{p:e}_{open} \mid D_{\text{close}}(b) \in d(r) \}$ is the set of bugs for a release $r\in R_p$ that were opened in an earlier release and that are closed in $r$.

    \item $B_\text{survive}^p(r) = \bigcup_{e \in earlier(r)} \{ b \in B^{p:e}_\text{open} \mid D_\text{close}(b) \geq date(r) \}$ is the set of bugs for a release $r\in R_p$ that are opened in an earlier release and not yet closed in $r$.

\end{itemize}

Using these notations, for $RQ_1$ we define \textit{fixing ratio} and \textit{fixing duration} of a project $p$ as follows. The \textit{fixing ratio} is the ratio of fixed over open bugs:

\begin{equation} \label{eq:fix-ratio}
    \textit{fixingRatio}(p) = \frac{|B_{fix}^p|}{|B_{open}^p|}
\end{equation}

The \textit{fixing duration} is the average fixing duration of all fixed bugs in a project. As the size of projects can vary, we normalize the impact of project size by dividing the total duration of fixed bugs by the number of fixed bugs in the project:

\begin{equation} \label{eq:fix-dur}
    \textit{fixingDuration}(p) = \newline \frac{\sum_{r \in R_p}\sum_{b\in B_{fix}^{p:r}} D_{close}(i) - D_{open}(i)}{|B_{fix}^p|}
\end{equation}

For $RQ_2$ we define the bug survival ratio of a project $p$ as the number of surviving bugs ($B_{survive}^p$) over the closed bugs, where bugs that are created and closed in the same release do not count towards the number of closed bugs of that release. Since such bugs have not survived over any release, they should not be considered in the survival ratio:

\begin{equation} \label{eq:suv_ratio}
\textit{survivalRatio}(p) = \frac{\sum_{r \in R_p}|B_\text{survive}(r)|}{\sum_{r \in R_p}|B_\text{newClosed}(r)|}
\end{equation}
    
For $RQ_3$ we define two bug triaging metrics. The \textit{triaging ratio} for a project $p$ is defined as the number of triaged bugs for every release of $p$, divided by the number of open bugs for every release of $p$:

\begin{equation} \label{eq:fix-ratio}
\textit{triagingRatio}(p) = \frac{|B_{triage}^p|}{|B_{open}^p|}
\end{equation}

The \textit{triaging duration} is defined as the average triaging duration of the triaged bugs in a project divided by the total number of triaged bugs in the project:

\begin{equation} \label{eq:triage-dur}
    \resizebox{1\hsize}{!}{
 $\textit{triagingDuration}(p) = \frac{\sum_{r\in R_p}\sum_{b \in B_{triage}^{p:r}} D_{triage}(b)-D_{open}(b)}{|B_{triage}^{p}|}$
}
\end{equation}
$RQ_4$ investigates how the variation in release cycle duration within a project affects the bug handling process. It uses the same metrics as $RQ_1$, namely fixing ratio and fixing duration. 
We use the interquartile range (IQR) instead of standard deviation for measuring the amount of variation~\cite{van2005statistics} as the distributions are very skewed and not normally distributed; the null hypothesis of the D'Agostino-Pearson test~\cite{normalitytest1,normalitytest2} that normality is met is rejected with p$<0.001$.
We answer $RQ_4$ by evaluating the effect of IQR on the bug fixing ratio and bug fixing duration.


\subsection{Gandhi-Washington Method}
The Gandhi-Washington Method (GWM) aims to analyze the impact of recurring events in software projects. To investigate the effects of event sequences, called treatments, on the dependent variable, called outcome, GWM uses three stages: \textbf{encoding}, \textbf{categorization} and \textbf{synthesis}.

The \textbf{encoding} stage receives the different event sequences together with their respective outcome value. The analyst encodes each event into a character of an \textit{alphabet}, allowing to represent the event sequences as strings in this alphabet. 
In our work, each encoded event corresponds to a single project release, and the character used to encode the event reflects the release duration. For example, a project with three consecutive releases with a very short duration of only 3 days each will be encoded as AAA; the chosen duration encodings will be detailed later in this section. The outcome value attached to each project will be different for each research question. For example, the outcome value for $RQ_2$ will be the project's bug $survivalRatio$. 

The \textbf{categorization} stage receives the encoded sequences of events with their outcome values. The first step consists of translating each encoding sequence into a regular expression. For example, AABBB and AAAABBBBB are both represented by the same regular expression A*B*.
The second step is to group the outcome values for which the regular expressions are equal. For example, AABBB and AAAABBBBB belong to the same group since they both correspond to the same regular expression A*B*. In the context of our work, this means that we will group projects with release sequences that are similar in terms of encoded release durations. 

The \textbf{synthesis} stage builds a hierarchy of regular expressions where items at the bottom of the hierarchy represent the most specific regular expressions while at the top we find the most general regular expressions. The synthesis iteratively applies the Mann-Whitney test~\cite{birnbaum1956use} to the outcome values of two regular expressions to determine if they are statistically different with significance level $\alpha=0.05$ after having applied a Bonferroni error correction to adjust for multiple comparisons.
If the two regular expressions are not statistically different, they are merged. When there are only regular expressions left with statistically significant different outcome variables, the testing stops. The remaining regular expressions indicate which sequences of events have an impact on the outcome variable.
In our work, the outcome of the synthesis step will reveal if the release cycle durations impact the bug handling activity.

GWM will be used to answer $RQ_1$ to $RQ_3$. Each sequence of events corresponds to the successive release durations for a single GitHub repository. Each release is represented as an integer encoding the {\em days until the subsequent release}, where releases within two days of each other will be considered the same release~\cite{joshi2019rapidrelease}. These integers are grouped into release duration ranges as follows:
{\bf A}: 2-5 days; {\bf B}: 6-20 days; {\bf C}: 21-35 days; {\bf D}: 36-365 days; {\bf E}: $\geq 366$ days until subsequent release.



\begin{figure}[!h]
    \centering
    \includegraphics[width=0.7\columnwidth]{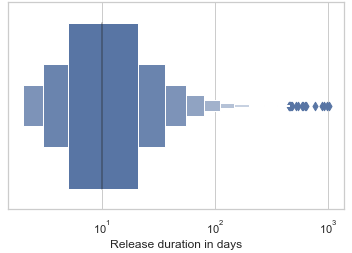}
    \caption{Boxenplot of the release durations distribution (in days) of all releases of all 420 investigated projects.}
    \label{fig:rel-dur-all-projects}
\end{figure}

This grouping is inspired by Joshi \etal~\cite{joshi2019rapidrelease} who defined three release cycle groups: $\leq 5$ days, between 6 and 35 days, and $\geq 36$ days. 
Given that 69\% (44,303 out of 63,956 releases) of the release cycle durations in our dataset belong to the [6..35] group, we
decided to split it into [6..20] (group {\bf B}) and [21..35] (group {\bf C}). This will allow us to differentiate between different types of short release cycles more carefully.
Similarly, we separated out group {\bf E} of $\geq 366$ days until subsequent release, as such release cycles are extreme outliers (0.1\% of the total releases, \ie 96 out of 63956). Figure \ref{fig:rel-dur-all-projects} shows the release duration distribution of all releases in all 420 projects. With a median release duration of 10 days it can be noted the releases are rapid releases and the majority of the releases falling within groups {\bf B} and {\bf C}.


The type of outcome value accompanying every release sequence will differ per research question. For $RQ_1$, GWM will be applied twice, with as outcome values the fixing ratio and the fixing duration, respectively. For $RQ_2$, the outcome value is the survival ratio. For $RQ_3$, GWM will be applied twice, with the triaging ratio and the triaging duration as outcome values.

%% file: sections/results.tex

\begin{table*}[h]
\centering
\resizebox{0.75\hsize}{!}{
\begin{tabular}{c|cccccccc}
group & {\bf Setting1} & Setting2 & Setting3 & Setting4 & Setting5 & Setting6 & Setting7 & Setting8\\\hline
A & {\bf 2-5} & 2-10 & 2-10 & 2-5 & 2-5 & 2-15 & 2-15 & 2-15\\
B & {\bf 6-20} & 11-30 & 11-20 & 6-35 & 6-25 & 16-35 & 16-35 & 16-35\\
C & {\bf 21-35} & 31-50 & 21-30 & 36+ & 26-50 & 36-75 & 36-75 & 36+\\
D & {\bf 36-365} & 51+ & 31+ & - & 51-100 & 76-365 & 76+ & -\\
E & {\bf 366+} & - & - & - & 101+ & 365+ & - & -
\end{tabular}
}
\caption{Different encodings for release cycle duration in the GWM tool. Cell contents indicate the range of release duration (lower to upper day limits) of the encoding w.r.t. the character in the first column of the respective row, while 
\textit{
``-"} indicates that this character is not used in the encoding.
\label{tab:settings-encoding}
}
\end{table*}



\textbf{$RQ_1$ \textit{How does release cycle duration affect the bug fixing process?}}\\
To analyze the bug fixing process, we will use GWM to answer two sub-questions:
$RQ_1^1$ \textit{How does release cycle duration affect \textbf{bug fixing duration}}? and $RQ_1^2$ \textit{How does release cycle duration affect the \textbf{bug fixing ratio}?} 
To do so, we use the encoding of release cycle durations presented in \sect{sec:methodology}.
We evaluated different encodings, presented in Table~\ref{tab:settings-encoding}, to ensure that the findings do not solely depend on the encoding chosen.
All settings were tested to find the optimal encoding to be used in our experiments. Since we did not observe any difference in the results, we decided to stick to the default  \textit{Setting1} of \sect{sec:methodology}.


\begin{figure*}[h]
    \centering{
    \includegraphics[width=.8\textwidth]{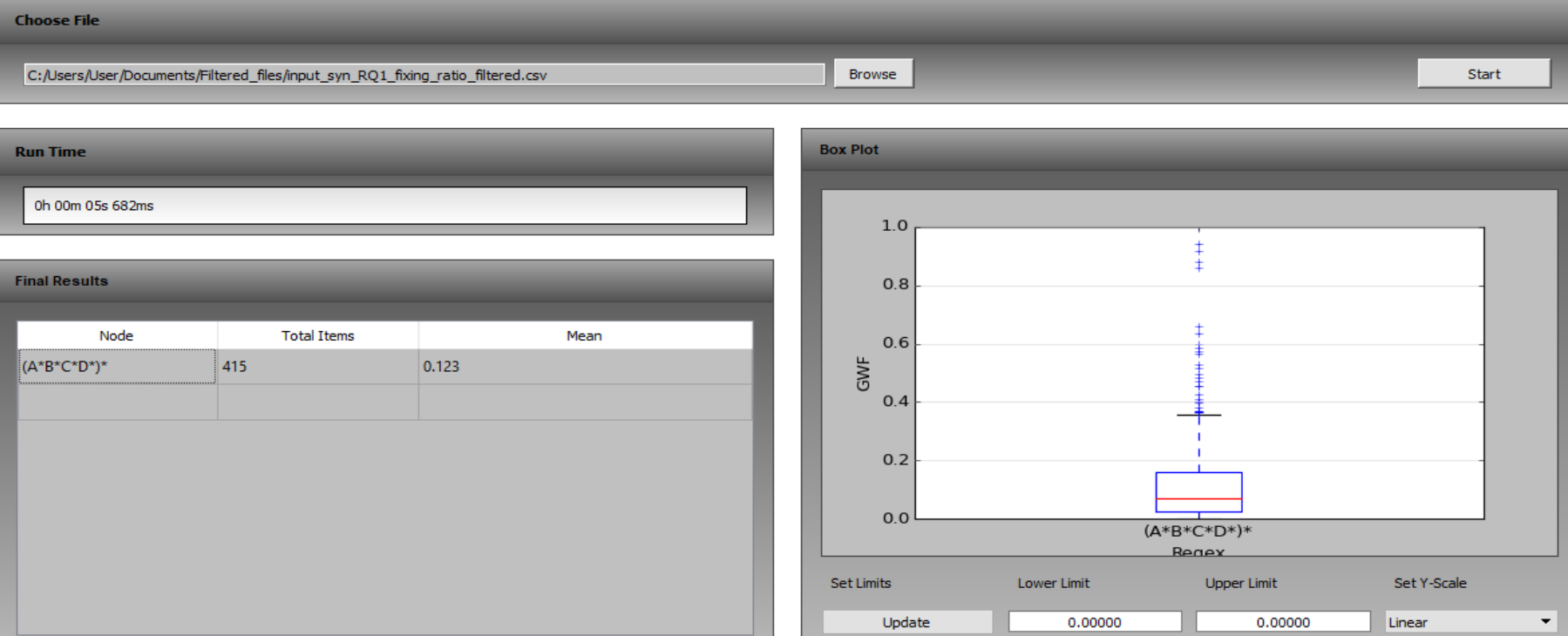}}  \label{fig:resultRQ1.1}%
    \caption{Figure showing the GWM tool and the results for $RQ_1^1$ for \textbf{Setting1} listed in Table \ref{tab:settings-encoding}}%
    \label{fig:rq1}%
\end{figure*}

For $RQ_1^1$ we used \textit{average fixing duration} as outcome value of GWM for each project, and for $RQ_1^2$ we used \textit{fixing ratio} as outcome value.
\fig{fig:rq1} presents the GWM results for $RQ_1^1$ and shows the tool in action.
The bottom left panel labeled final results contains the statistically significant regular expressions with the average of each regular expression's outcome value, and the right panel shows a boxplot with the distributions of the outcome values of the different statistically significant regular expressions.

For both sub-questions, the results contains only a single statistically significant regular expression \textit{(A*B*C*D*)*} covering all possible sequences. This implies that there is no difference in fixing duration or fixing ratio between projects with different release cycle durations. For example, the results for short releases (group \textbf{A} with durations of 2 to 5 days) are not different from those for very long releases (durations in group \textbf{D} between 35 and 365 days) or even an arbitrary mixture of release durations. This goes against the intuition that developers would treat bugs differently in release sequences of very short duration compared to release sequences of very long duration. A possible explanation might be that, while developers will handle less bugs in shorter releases, they do not handle individual bugs differently. 


%



\begin{mybox}
Different release cycle durations do not statistically significantly impact the bug fixing duration/fixing ratio.
\end{mybox}

\textbf{$RQ_2$ \textit{How does release cycle duration affect bug survival over multiple releases?}}

For $RQ_2$ we apply GWM using the bug \textit{survival ratio} of a project as outcome value. 
Again, the result contains only a single statistically significant regular expression \textit{(A*B*C*D*)*} that covers all sequences of release cycle durations.
This implies that the bug surviving ratio does not differ between projects with different release durations. Again, this goes against the intuition that bugs might be postponed to later releases if there is less time to fix bugs in very short releases. 
A possible explanation for why the observed results do not confirm this intuition is that, when developers know there is little time for each release, they plan the release such that bug handling still receives enough resources  to avoid having very persistent bugs. 

\begin{mybox}
Difference release cycle durations do not statistically significantly impact the bug survival ratio. 

\end{mybox}


\textbf{$RQ_3$ \textit{How does release cycle duration affect bug triaging?}}

To analyse the bug triaging process we will study two subquestions concerning the triaging duration and ratio: $RQ_3^1$ \textit{How does release cycle duration affect \textbf{bug triaging duration}?}
and $RQ_3^2$ \textit{How does release cycle duration affect the \textbf{bug triaging ratio}?}
We use GWM to answer both subquestions, using as outcome values the triaging duration and triaging ratio for each project, respectively.
For both subquestions, the results again indicate there is only one statistically significant regular expression (A*B*C*D*)*.
This implies that there is no difference in triaging duration or triaging ratio between projects with different release cycle durations.


With triaging taking the same amount of time for any duration of a release, we observed that short duration releases have their bugs triaged later into the release, ratio wise, compared to the longer duration release. Additionally, at the same time the ratio of triaged bugs stays the same.

This is counter intuitive as we expect more bugs to be triaged and fixed near the end of a release due to the increased pressure closer to the release deadline. 





%


\begin{mybox}
Difference release cycle durations do not statistically significantly impact the bug triaging ratio/duration. 


\end{mybox}

\textbf{$RQ_4$ \textit{How do variations in the release cycle duration of a project affect the bug handling process?}}\\

\begin{figure}[h]
    \centering{
    \includegraphics[width=.75\columnwidth]{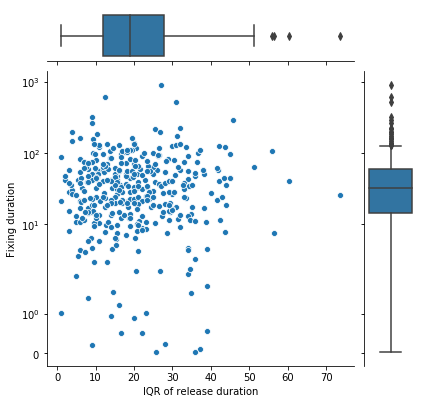}} 
    \caption{Comparing distributions of average fixing duration and IQR of release duration}
    \label{fig:IQR-fix-dur}
\end{figure}

\begin{figure}[h]
    \centering{
    \includegraphics[width=.75\columnwidth]{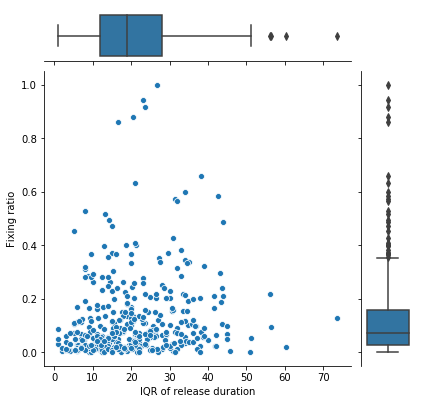}}
    \caption{Comparing distributions of bug fixing ratio and IQR of release duration.}
    \label{fig:IQR-fix-ratio}
\end{figure}

 We compute the IQR of the release durations for all projects in the dataset. Then, we find each project's average fixing duration and overall fixing ratio. \fig{fig:IQR-fix-dur} and \fig{fig:IQR-fix-ratio} visualise the distribution of the IQR versus the two metrics respectively.


We verify if the distribution of IQR values is normally distributed by applying the D'Agostino-Pearson normality test\cite{normalitytest1,normalitytest2} with null hypothesis that the data is normally distributed. Since $p=3.01e-23$ we reject the hypothesis and conclude that the distribution is not normally distributed. To evaluate whether the variation in release cycle duration is correlated to bug fixing, we use Spearman's rank correlation coefficient \cite{gautheir2001detecting}. The null hypotheses $H0_4^1$ and $H0_4^2$ state that there is no correlation between the variation in release cycle duration and the fixing ratio (respectively, fixing duration). 

We found no to a very low positive correlation between the variability of the release cycle duration and a project's fixing ratio or fixing duration; correlation values were \textit{0.021} for $H0_4^1$ and \textit{0.234} for $H0_4^2$. This means that we cannot reject the null hypotheses, since the variability in release cycle duration does not correlate well  the bug fixing ratio and fixing duration. 



\begin{mybox}
We do not observe a statistical correlation between variation in release cycle duration and bug fixing duration/ratio. 
\end{mybox}

None of the research questions revealed any statistically significant relation between different release cycle durations and the respective metrics. Thus, we have not been able to find any statistically significant correlation between the release engineering process and bug handling activity. 


%% file: sections/discussion.tex

Previous research \cite{abou2019longitudinal,khomh2015understanding,da2018impact, souza2014rapid} showed a difference in bug handling activity when comparing traditional to rapid release cycles. 
This lead to the hypothesis that different rapid release cycles may reveal a difference in bug handling activity. 
However, our results did not confirm this hypothesis.
A possible reason could be that the projects in the RapidRelease dataset are not as popular or mature as large and mature projects (such as Firefox or Eclipse) that have been studied in previous research.
In addition to this, there is anecdotal evidence that compact, cross-functional and efficient teams are needed to develop efficiently in the presence of rapid releases.\footnote{\url{https://techbeacon.com/devops/doing-continuous-delivery-focus-first-reducing-release-cycle-times}} Hence, one can assume that rapid release cycles are mostly beneficial for mature projects and development teams. 
Once the release cycle starts to become too short, however, the added value it brings would be wasted, since it may lead to a reduced code quality (due to time pressure), resulting in additional bug handling activity.\footnote{\url{https://www.overops.com/blog/fast-release-cycles-wasting-developer-time/}} This could be another possible explanation for the lack of difference in bug handling between the shorter and longer rapid release cycles. 

A different observation that grabs attention is the lack of difference between the variability in releases and the fixing ratio/ fixing duration. 
For projects with a fixed release plan, one would expect better organization and clear development goals, e.g., which functionality to implement for each release, compared to projects with variable release duration. However, the variability does not seem to affect the bug fixing activity (see $RQ_4$). We argue that neither a loose release schedule necessarily negatively affects the bug handling activity nor a strict release schedule necessarily improves it.

We rationalize that developers pick a release duration to support providing the best quality software to users. This includes providing new functionality regularly and ensuring the software's quality with respect to bugs. This means developers would pick a release duration which also helps bug handling activity. 


Thus, we hypothesize that a fixed release schedule combined with other factors, e.g. community contribution, may have a positive effect on bug handling activity. However, none of our obtained results seem to point in this direction. $RQ_4$ shows that variability in the release cycle duration does not affect the bug handling activity, suggesting that focusing extra resources on obtaining a fixed release schedule does not necessarily improve the project's bug handling activity.

It can be noted that changing the encodings for the release cycle duration groups does not affect the results for the different research questions. If one of the encodings resulted in more than one statistically significant regular expression, it would indicate the reasoning behind the default release cycle groups encoding is wrong. However, for all settings, there is only one statistically significant regular expression for each research question. This strengthens the observation that there is no correlation between the rapid release cycle duration and the respective bug handling performance metrics. 


Overall, researchers investigating factors which may affect bug handling activity between rapid releases, or building frameworks of actionable factors that affect bug handling activity, must be aware that while different works highlight that the release cycle duration affects bug handling activity when moving to more rapid release cycles, we have found no such correlation between different rapid release cycles and bug handling activity in the dataset of projects that were studied.

%% file: sections/threaths-validity.tex
Following the structure presented by \textit{Runeson et al.}~\cite{runeson2009guidelines}, we address the threats that can affect the study's validity and highlight how we have tried to mitigate these threats. 

A threat to the construct validity of our work concerns identifying bugs among the issues in the GitHub repositories. Not all GitHub repository issues represent bugs, so we classified labels assigned to issues into \textit{bugs} and \textit{non-bugs}, marking issues as bugs if they had at least one bug label. This may have caused some bugs to be labeled as \textit{non-bugs} and vice-versa. We have mitigated this risk by minimizing the bias stemming from manual labeling: at least two of the authors labeled each issue, and disagreements were discussed among all authors until a consensus was reached.

A threat to the internal validity of the work is the fact that bugs can be reopened when it turns out that the problem was not adequately fixed. We opted to take the last fixing date in the presence of multiple fixes as fixes prior to the last one were not satisfactory.
Another threat to internal validity stems from considering the link between bugs and commit/pull requests to consider a bug as fixed. Although not all fixed bugs are linked to the fixing commits~\cite{bird2009fair}, this is the recommended behavior in GitHub~\cite{liu2016comparative} and cannot be further mitigated.
Finally, the threat that not all issues are labelled on GitHub is addressed by filtering out repositories with less than 40\% labelled issues.

Regarding external validity, we cannot generalize our results to repositories having non-rapid release cycles. Projects with longer release cycles are likely to reveal different behavior in their bug handling characteristics. Additionally, we only consider projects using GitHub issues instead of dedicated bug tracking platforms. As different bug tracking platforms might affect how developers approach the bug handling activity, the results cannot be generalized to projects using bug tracking platforms other than GitHub. Finally, we cannot generalize our findings to projects with different characteristics than the ones in the RapidRelease dataset w.r.t. attributes like community size, maturity, etc.

%% file: sections/related-work.tex


da Costa \etal~\cite{da2018impact} showed that rapidly releasing projects introduces delays in the integration of fixed issues. They performed a comparative study of 72,114 issue reports from the open source project Firefox before and after its switch to a rapid release cycle. They found that issues are fixed faster in rapid releases but, surprisingly, rapid releases take longer than traditional releases to integrate fixed issues. 
Khomh \etal~\cite{khomh2015understanding} also investigated Firefox and observed that faster release cycles improve software quality. The authors investigated crash rates, median uptime, and the proportion of post-release bugs of releases with a short cycle vs releases with a longer ``traditional" release cycle. They found that bugs are fixed faster, and there are not more post-release bugs in releases with a rapid cycle compared to releases with a traditional cycle length. However, the users of the software run into critical execution problems with the program sooner. 

Abou Khalil \etal~\cite{aboukhaliljss2020,abou2019longitudinal}
analyzed various aspects of bug fixing in the open source project Eclipse. They investigated how the bug handling process differs between the pre-release and post-release periods after Eclipse switched from yearly to quarterly release cycles. They observed that there is no difference in bug fixing before or after a release, except for triaging that is faster before the release. Additionally, they observed that bugs were triaged and fixed faster after the switch to the quarterly release cycle duration. They obtained feedback from five Eclipse Core maintainers who confirmed their findings.

To determine whether the release cycle duration affects the app rating, Maleknaz \etal~\cite{nayebi2019mining} analyzed 6,003 mobile apps through the GWM method. 
They found seven unique release sequences that significantly affect the app rating. Also, they found that apps with consecutive long release cycles, followed by consecutive short release cycles have the highest median app rating. In our paper, we used GWM to analyze the effect of release cycle duration on bug handling activity.

All the above studies are valuable in understanding how rapid release impacted the bug handling process. In contrast, we studied how the varying durations of rapid release cycles impact the bug handling activity.

%% file: sections/conclusion.tex


In this work, we performed an empirical study on 420 projects with a rapid release cycle. We investigated the effect of the rapid release cycle duration and the variation of the release cycle duration on the bug handing activity. The bug handling activity was measured in terms of fixing duration, fixing ratio, bug survival over releases, and bug triaging duration/ratio. 
We used the Gandhi-Washington algorithm to evaluate whether the effects of the release cycle duration on the bug handling activity are statistically significant.

Contrary to our expectations, our findings did not reveal any statistically significant impact of the release cycle duration or its variability on the bug handling activity. These findings suggest that differences within rapid release cycle duration do not affect the bug handling activity. 

Future research could enlarge the dataset of projects investigated. Different bug trackers than GitHub can be investigated to determine if the observed results are specific to the bug tracking platform. Additionally, the dataset of the GitHub projects investigated can be diversified. Adding larger, more mature projects should be the priority in diversifying the GitHub projects. 